\documentclass[
 reprint,
 superscriptaddress,
 amsmath,
 amssymb,
 aps,
 prl,
]{revtex4-2}

\usepackage{graphicx}
\usepackage{dcolumn}
\usepackage{bm}
\usepackage[colorlinks=true, hyperindex, breaklinks, urlcolor=blue, linkcolor=black, citecolor=blue]{hyperref}
\usepackage[english]{babel}
\usepackage{amsmath,graphicx,textcomp}
\usepackage{xcolor}

\definecolor{db}{RGB}{1, 33, 105}
\hypersetup{citecolor=db,urlcolor=db}

\newcommand{\tmop}[1]{\ensuremath{\operatorname{#1}}}

\newcommand{\ece}{Department of Electrical and Computer Engineering, Princeton University, Princeton, New Jersey 08544, USA}

\newcommand{\duke}{Department of Electrical and Computer Engineering, Duke University, Durham, North Carolina 27708, USA}

\begin{document}

\preprint{APS/123-QED}

\title{A scaled local gate controller for optically addressed qubits}

\author{Bichen Zhang}
 \thanks{These authors contributed equally to this work.}
 \affiliation{\ece}

\author{Pai Peng}
\thanks{These authors contributed equally to this work.}
\affiliation{\ece}

\author{Aditya Paul}
\affiliation{\duke}

\author{Jeff D. Thompson}
\email{jdthompson@princeton.edu}
\affiliation{\ece}

\begin{abstract}
Scalable classical controllers are a key component of future fault-tolerant quantum computers. Neutral atom quantum computers leverage commercially available optoelectronic devices for generating large-scale tweezer arrays and performing parallel readout, but implementing massively parallel, locally-addressed gate operations is an open challenge. In this work, we demonstrate an optical modulator system based on off-the-shelf components, which can generate a two-dimensional array of over 10,000 focused spots with uniform frequency and amplitude, and switching them on and off individually in arbitrary configurations at rates of up to 43 kHz. Through careful control of aberrations, the modulator achieves an extinction ratio of 46 dB, and nearest-neighbor crosstalk of $-44$ dB with a beam spacing of 4.6 waists. The underlying components can operate at wavelengths from the UV to the NIR, and sustain high laser intensities. This approach is suitable for local addressing of gates with low cross-talk error rates in any optically addressed qubit platform, including neutral atoms, trapped ions, or solid-state atomic defects.
\end{abstract}

\maketitle
\onecolumngrid

Neutral atoms in tweezer arrays have recently emerged as a promising platform for quantum science and technology, with applications to quantum computing~\cite{bluvstein2022quantum, graham2022multiqubit,ma2023high}, many-body simulations~\cite{browaeys2020many,bluvstein2021controlling,semeghini2021topologically}, and metrology~\cite{madjarov2019atomic,young2020half}. One of the key advantages of neutral atom arrays is that near-term scalability is enabled by leveraging commercially available optoelectronic devices. For example, CMOS and CCD cameras have enabled parallel readout of hundreds of qubits~\cite{ebadi2021quantum, schymik2022equalization, huft2022simple}, and liquid-crystal-on-silicon spatial light modulators (LCOS-SLMs) and acousto-optic deflectors (AODs) allow the creation of arrays of thousands of tweezer traps, with dynamic reconfiguration~\cite{barredo2016assembler, endres2016atom,  kim2016situ}. The resolution of these devices is approximately 1 megapixel, which is compatible with scaling beyond ten thousand tweezers.

However, scalable local control of gate operations is an outstanding challenge for this platform~\cite{graham2022multiqubit,burgers2022controlling}. This challenge is shared by other optically addressed qubits such as trapped ions~\cite{debnath2016demonstsration,pogorelov2021compact,kranzl2022controlling} and solid state defects. An ideal controller must be able to generate uniform arrays of focused spots, and quickly switch between arbitrary illumination patterns with high on/off contrast and low crosstalk between closely spaced sites. Furthermore, the controller must be able to operate at application-dependent wavelengths and intensities, which range from UV to IR and often require milliwatts or more power per site.

Several approaches to overcoming this challenge have been pursued. For example, multi-channel acousto-optic modulators (MC-AOM) \cite{debnath2016demonstsration} and AOM arrays \cite{binai2023guided} have been demonstrated in small arrays of up to 32 channels, but scaling to larger numbers requires assembly of many discrete optical and electronic components. Acousto-optic deflectors (AODs) \cite{graham2022multiqubit,pogorelov2021compact,kranzl2022controlling, li2023low} have been used to demonstrate individual addressing of around 50 qubits in both 1D and 2D. However, they are limited to row or column addressing when generating multiple spots in parallel, and the contrast can be limited by nonlinear intermodulation effects. LCOS-SLMs can generate arbitrary illumination patterns, but have refresh rates of 60-120 Hz,  much slower than the intrinsic gate times of atomic qubits. Faster spatial light control can be realized with digital micromirror devices (DMDs)~\cite{zupancic2016ultra}, but is inefficient when used to generate sparse spot arrays~\cite{shih2021reprogrammable}. Very recent work has demonstrated gate controllers using novel photonic devices, including photonic integrated chips (PICs)~\cite{menssen2022scalable, christen2022integrated}, and MEMS-based beam steering systems (MEMS-BSS)~\cite{knoernschild2010independent, wang2020high}. While PICs are promising for realizing very high switching speeds, they have not yet demonstrated a channel count beyond a few tens, or characterized crosstalk between closely spaced spots. MEMS-BSS is promising for low crosstalk and extremely high contrast, but has not demonstrated parallel control of more than two beams. A recent proposal combining an AOD with a segmented LCOS-SLM overcomes some of these limitations~\cite{graham2023multiscale, kim2021optical}.

In this work, we demonstrate a new approach to generating large-scale arrays of individually controlled laser beams for local gate operations. It is based on a combination of three commercially available modulators, used in series to implement separate functions on separate timescales. First, an AOM is used to generate a pulse of light with the desired frequency and waveform in a single spatial mode, with nanosecond-scale temporal resolution. Then, an LCOS-SLM diffracts the pulse into an array of secondary beams at fixed positions, corresponding to the qubit locations. Finally, a DMD placed in an image plane is used to selectively shutter the secondary beams, which determines which subset of the qubits are ultimately illuminated; the DMD can be reconfigured to illuminate different subsets of qubits every $23\,\mu$s. We achieve an extremely high extinction ratio by operating the DMD as a diffraction grating, with a locally switchable blaze angle. The key challenge is controlling aberrations arising from diffracting tightly focused beams with the DMD, which results in site-to-site crosstalk and limits the spot size uniformity. We analytically design and implement a correction system consisting of a telescope and a ruled grating. With this approach, we demonstrate an array of 10,000 beams separated by $4.6 w_0$ (where $w_0$ is the $1/e^2$ radius), with 10\% uniformity in the beam waist and 1.6\% uniformity in the intensity across the array. The average on/off contrast of each site is 46 dB, and the average crosstalk between nearest-neighbor sites is $-44$ dB. The combined diffraction efficiency of the SLM and DMD is 0.37, when all 10,000 spots are switched on.

This modulator is suitable for controlling parallel gate operations in large-scale neutral atom arrays. It can be employed to focus gate beams directly ({\emph i.e.}, drive atomic transitions) or to apply local light shifts~\cite{labuhn2014single, deleseleuc2017optical, levine2018high}, which is a particularly robust approach for nuclear spin qubits in alkaline earth atoms~\cite{burgers2022controlling, ma2022universal, ma2023high, lis2023mid, huie2023repetitive, norcia2023mid}. This approach can also be used for other optically addressed qubits, including trapped ions and solid-state defects~\cite{Rovny2022Nanoscale}, and other applications including quantum simulation and metrology~\cite{madjarov2019atomic,young2020half}.

\begin{figure*}[!htp]
    \centering
    \includegraphics[width=180 mm]{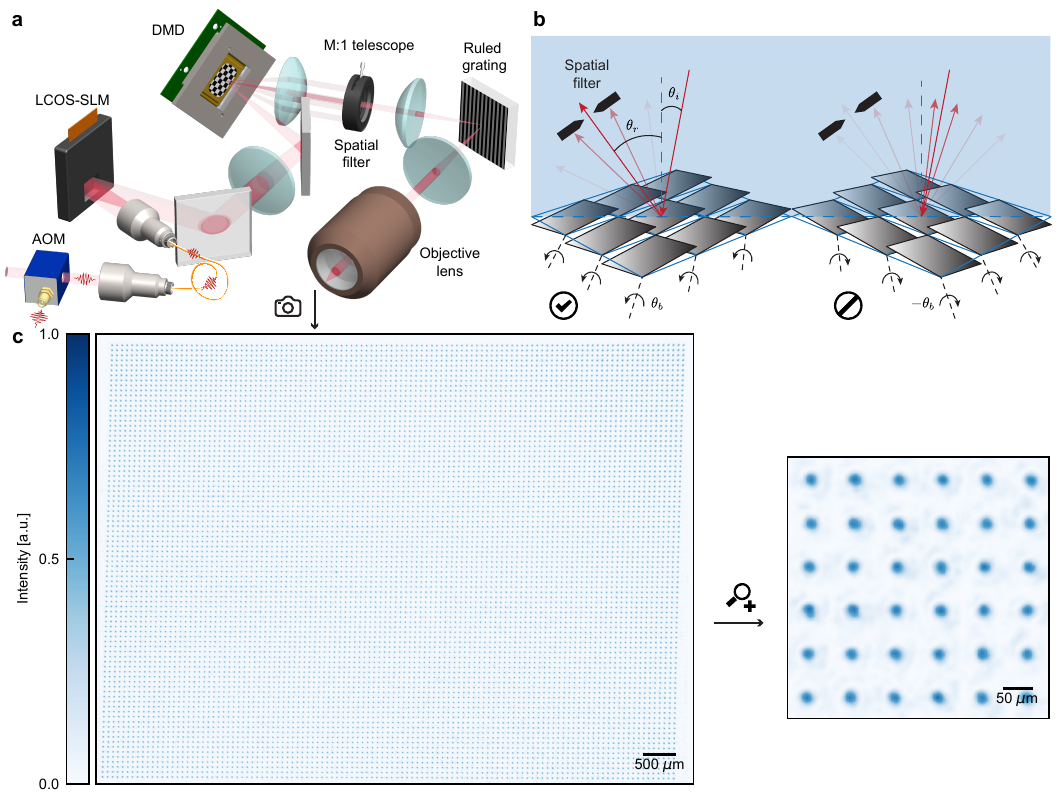}
    \caption{\textbf{Scaled local gate controller approach} (a) Schematic diagram of the controller, consisting of an AOM generating fast laser pulses, an LCOS-SLM splitting the pulses into 10,000 secondary spots, and a DMD switching arbitrary subsets of spots on and off dynamically. We use a $M=1:1$ telescope and a 600 groove/mm ruled grating to fix the aberrations caused by the tilted axes of the imaging system with respect to the DMD plane. (b) Illustration of the DMD switching on (off) a secondary spot by satisfying (violating) the blazing condition. The angle of incidence ($\theta_i$), the angle of reflection ($\theta_r$), and the blaze angle of DMD ($\theta_b$) are marked in the figure. A spatial filter is used to block all diffraction orders except the desired one. (c) Image of a $115\times87$ array with spot spacing at $4.6w_0$, with all DMD mirrors switched on.}
    \label{fig:intro}
\end{figure*}

The basic setup of the controller is shown in Fig. \ref{fig:intro}a. First, an AOM generates pulses with nanosecond-scale timing resolution. Then, an LCOS-SLM splits the single, primary laser beam into 10,000 secondary beams of an arbitrary geometry by imprinting the phase in the Fourier plane~\cite{nogrette2014, kim2016situ, kim2019largescale}. The geometry of the SLM beams can be reconfigured but is assumed to be static over the course of executing one circuit, as the LCOS-SLM frame rate is only 60 Hz. Finally, a DMD acts as an optical switch array to rapidly activate or deactivate specific subsets of the secondary beams, to control which secondary spots are ultimately transmitted to the qubits. The DMD (Texas Instruments DLP7000) is an array of 768 $\times$ 1024 micromirrors with pitch $a=13.68\,\mu$m that can be switched between two, fixed tilt angles, $\theta_b = \pm 12.35^\circ$. Under coherent illumination from a laser, the DMD acts as a diffraction grating. We choose the angle of incidence to satisfy a blazing condition when the mirrors are in the $+ 12.35^\circ$ state, such that the reflected light is concentrated into a single outgoing diffracted order. When the mirrors are in the opposite state, the reflected light is spread out across many orders. Provided the beam waist on the DMD is larger than the mirror pitch, the angular separation between the diffraction orders is larger than the divergence of the focused beams, allowing the unwanted orders to be blocked by a spatial filter (Fig. \ref{fig:intro}b). In this work, the beam waist on the DMD is $20.4\,\mu\textrm{m} = 1.49 a$.

The DMD parameters dictate that the angle of incidence and reflection are relatively large, which makes it difficult to construct a single imaging system to focus and recollimate the spot array across the entire DMD aperture without aberrations (in our setup, $\theta_i = 11.1^\circ$ and $\theta_r = 35.8^\circ$). As an alternative to enable the use of off-the-shelf optics, we implement separate imaging systems for the incident and outgoing beams, with tilted optical axes aligned to $\theta_i$ and $\theta_r$, respectively. A consequence of this choice is that the DMD does not lie in the focal plane of the imaging system, which results in a position-dependent defocus and astigmatism across the DMD aperture. However, these aberrations can be corrected by using a second, compensating diffraction grating after the DMD (Fig. \ref{fig:intro}a, see Supplementary Information). With this approach, we can generate 10,005-spot arrays ($87\times115$) with a separation of $4.6w_0$ (Fig. \ref{fig:intro}c). This corresponds to approximately 210,000 diffraction-limited modes, essentially saturating the capacity of the DMD. In contrast, without the compensation grating, less than $1/3$ of the DMD resolution is usable.

\begin{figure*}[!htp]
    \centering
    \includegraphics[width=180mm]{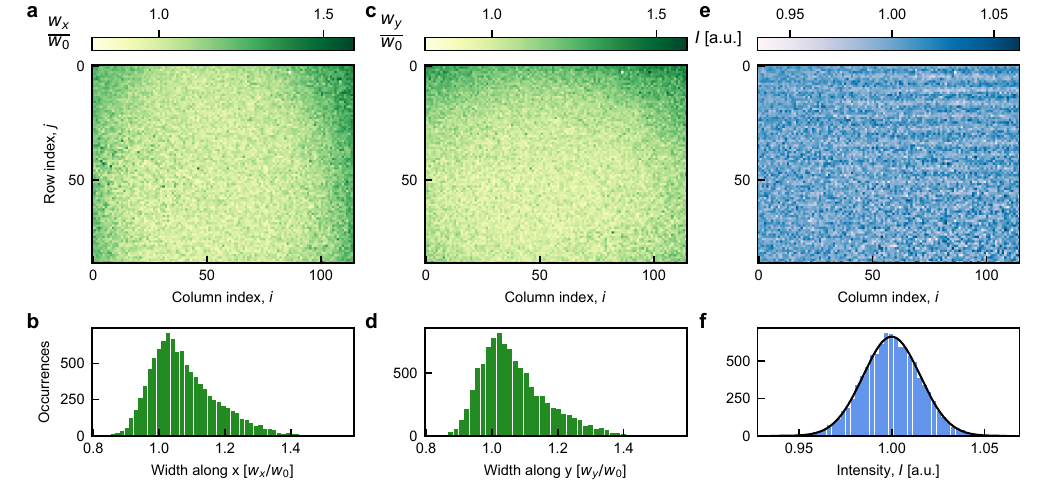}
    \caption{\textbf{Uniformity across the array} (a) Map of the horizontal beam waist $w_x$ ($1/e^2$ radius) across the array, normalized to the beam waist in the center of the array, $w_0 = 17.2\,\mu$m. (b) Histogram of $w_x$ values. The standard deviation is $10.0\%$. (c) Map of the vertical beam waist, $w_y$. (d) Histogram of $w_y$ values. The standard deviation is $10.2\%$. (e) Map of the spot intensity across the array, normalized to the mean intensity. The horizontal stripes result from the interplay of the 120 Hz LCOS-SLM flicker and the 23 kHz line rate of the CMOS camera rolling shutter. (f) Histogram of spot intensities, with standard deviation $1.6\%$. All parameters are obtained by fits to the intensity recorded on a camera, using a 2D Gaussian function.}
    \label{fig:uniform}
\end{figure*}

To characterize the performance, we first study the uniformity of the generated array. In these experiments, we focus the spot array directly onto a camera with an $f=200$ mm achromatic doublet, resulting in a beam waist of $w_0 = 17.2\,\mu$m. Across the $10,005$-spot array, we find that the beam waist has a standard deviation of approximately $10\%$ (Fig. \ref{fig:uniform}a-d). We also characterize the spot intensity, after homogenizing the array with the LCOS-SLM using the weighted Gerchberg-Saxton algorithm~\cite{gerchberg1972practical, di2007computer, kim2019largescale}. We find a standard deviation of $1.6\%$ (Fig. \ref{fig:uniform}e, f). A significant fraction of the residual intensity non-uniformity arises from the 120 Hz flicker on the LCOS-SLM, which we have made visible in Fig. \ref{fig:uniform}e by choosing a synchronous camera frame rate. 

\begin{figure*}[!htp]
    \centering
    \includegraphics[width=180mm]{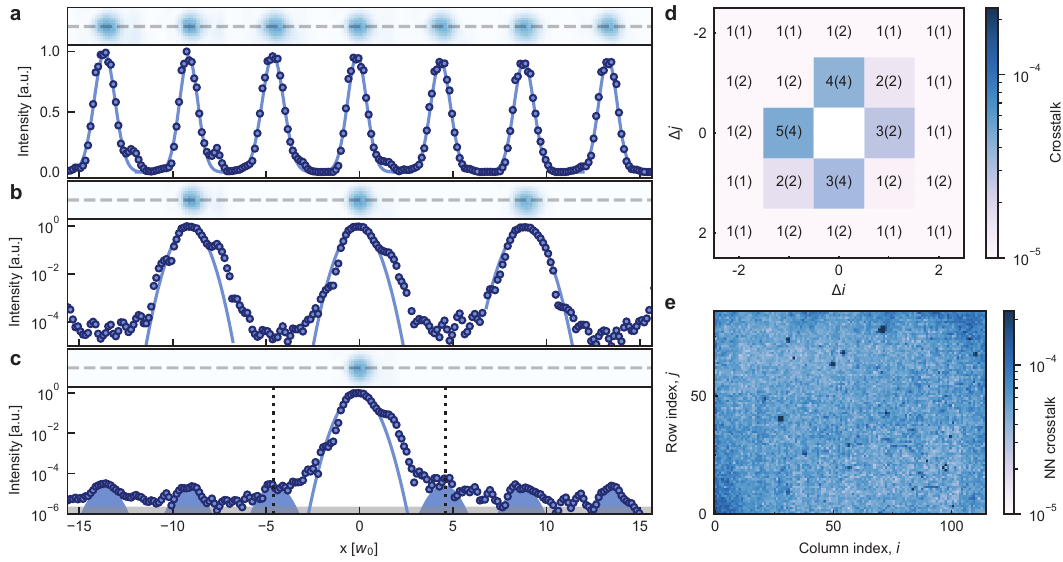}
    \caption{\textbf{Contrast and crosstalk characterization}. (a) Cross section of the intensity (normalized to the peak intensity) in a row with all spots switched on. (b) Cross section of the intensity when every other spot in a row is on. (c) Cross section of the intensity when a single spot is on. Sites far from the illuminated site have a residual intensity of $2.4\times10^{-5}$ or $-46$ dB (shown as blue-shaded region), while closer sites have additional intensity error from crosstalk (at a level of $3\times10^{-5}$ for the right adjacent site). The grey-shaded area represents the noise floor of the HDR images. (d) Average crosstalk on sites displaced by $(\Delta i, \Delta j)$ from an isolated illuminated site. The values labeled in the plot are scaled by $10^{5}$. (e) Map of the average nearest neighbor crosstalk across the entire array. The scattered areas of high crosstalk are caused by cosmetic defects on compensation grating.}
    \label{fig:crosstalk}
\end{figure*}

A crucial figure of merit for locally addressed gate operations is the contrast (\emph{i.e.}, on/off intensity ratio of a single site) and crosstalk (\emph{i.e.}, unintentional illumination of sites around a target site). In Figs. \ref{fig:crosstalk}a-c, we show line cuts of a single row of the array with all sites, every other site, and only one site illuminated. In all cases, the LCOS-SLM is generating the full 10,005-site pattern, and a subset of sites is selected using the DMD only. By taking a series of images with logarithmically spaced exposure times (\emph{i.e.} high dynamic range imaging), we are able to record the intensity with a dynamic range of approximately $10^6$. In images with sparse illumination such as Fig.~\ref{fig:crosstalk}c, the average intensity at sites far from an illuminated site is $2.4\times10^{-5}$ ($-46$ dB), which we define as the contrast of the modulator. Sites near the target site also experience crosstalk, which has an average value of $4\times10^{-5}$ ($-44$ dB) across the array. Remarkably, this performance is maintained across the entire array: Fig.~\ref{fig:crosstalk}e shows the nearest neighbor crosstalk when illuminating each of the 10,005 sites individually. Isolated spots with higher crosstalk result from blemishes in the compensation grating (a replicated epoxy grating), which are imaged onto the final array. Fig.~\ref{fig:crosstalk}d shows the average crosstalk as a function of displacement from the illuminated site, indicating a rapid decay with distance.

The high contrast results from the excellent extinction of light diffracted into the target order when the DMD mirrors are toggled. An ideal diffraction grating with the same parameters has a theoretical contrast of $5\times10^{-7}$, ($-63$ dB) for a plane-wave input, while a finite-difference time-domain (FDTD) model incorporating realistic imperfections (\emph{e.g.}, finite fill factor, rounded corners, and the hinge hole in the mirror center) predicts a contrast of $5.9\times10^{-6}$ or $-52$ dB (see Supplementary Information). This is close to the experimental value, despite the fact that each beam only spans several DMD pixels. On the other hand, we do not have a quantitative model for the crosstalk onto nearest neighbor sites, but have observed that it is extremely sensitive to aberrations arising from the fiber collimator, clipping of the beam on apertures, including the DMD, as well as the precise alignment of the setup.

\begin{figure}[!htp]
    \centering
    \includegraphics[width=90 mm]{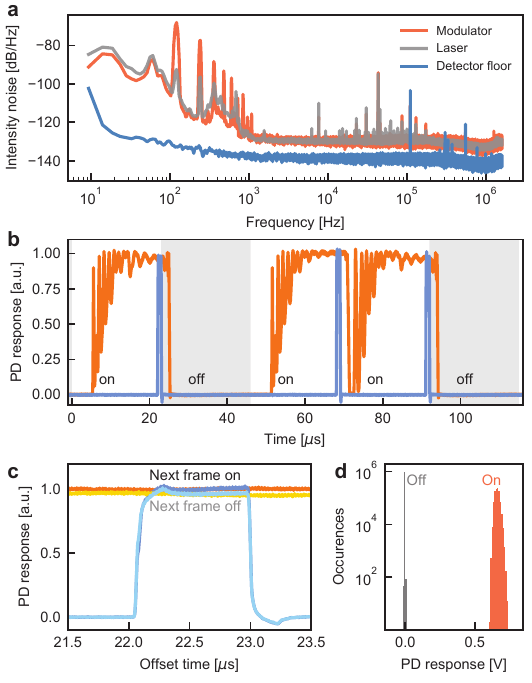}
    \caption{\textbf{Dynamical performance} (a) Power spectral density of the intensity noise in a single transmitted spot, under continuous-wave illumination (orange). The grey curve shows the intensity noise of the input laser, while the blue line shows the noise floor of the photodiode. (b) Intensity of a single spot while changing the DMD pattern at 43.5 kHz (frame time $T = 23~\mu$s). The orange (blue) curve shows the output intensity of a single spot when the AOM is always on (pulsed). (c) Overlaid traces of two frames. Dark-(light-) colored traces show the PD response of a frame whose next frame is on(off). (d) Histogram of photodiode response during a 2-million-flip pseudorandom bit sequence, showing no errors.}
    \label{fig:dynamics}
\end{figure}

Next, we characterize the dynamic performance of the system (Fig.~\ref{fig:dynamics}). In these measurements, we replace the camera with a photodiode, and configure the DMD to transmit only a single spot. The power spectral density under continuous illumination is shown in Fig.~\ref{fig:dynamics}a. Relative to the source laser, the transmitted beam has added intensity noise at low frequencies, particularly 120 Hz and its harmonics, which we attribute to flicker from the refresh rate of the LCOS-SLM. However, there is no measurable noise added at frequencies beyond 1 kHz (Fig. \ref{fig:dynamics}a). Next, we characterize the dynamics while switching the DMD mirrors (Fig. \ref{fig:dynamics}b).  With continuous laser illumination, strong transients are visible when DMD micromirrors are switched. However, the intensity is stable after the mirrors have settled. At the full frame rate of 43 kHz, microsecond-long pulses can be applied during the stable region, exhibiting a high stability with a pulse flatness characterized by less than 0.8\% intensity variation; longer stable windows can be achieved with slower frame rates. Unexpectedly, we observed that the intensity changes by about 3\% depending on the state of the mirrors in the {\em next} frame, which arises from a small electrostatic force on the mirror from the CMOS memory cell underneath, holding the next state of the mirror~\footnote{ViALUX, private communication}. Finally, we characterize the error rate using a pseudo-random sequence of 2 million frames at the full frame rate. We do not observe any errors, resulting in an upper bound of the bit error rate of $5\times10^{-7}$ (Fig. \ref{fig:dynamics}c). 

\begin{figure*}[!htp]
    \centering
    \includegraphics[width=115 mm]{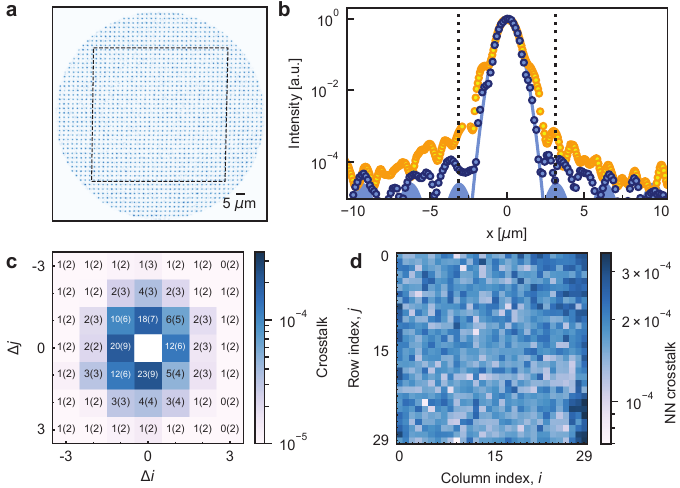}
    \caption{\textbf{Contrast and crosstalk characterization after focusing with a microscope objective} (a) Camera image of the array after focusing to a beam waist of $w_0 = 800$ nm. The field of view is limited by the objective lens used to image the array, and we only consider the central 30 $\times$ 30 sites (dashed box) where the imaging is free of distortion. (b) Cross-section of the intensity (normalized to the peak intensity) with a single spot illuminated before (yellow) and after (blue) aberration compensation using the LCOS-SLM. The dotted vertical lines show the position of the neighboring (off) sites. (c) Average crosstalk on sites displaced by $(\Delta i, \Delta j)$ from an isolated illuminated site. The values labeled in the plot are scaled by $10^{5}$. (d) Map of the nearest-neighbor crosstalk across the central $30\times30$ sites.}
    \label{fig:objective}
\end{figure*}

To model realistic experimental conditions in a neutral atom quantum computer, we replace the achromatic doublet focusing lens with a microscope objective (Olympus Plan Achromat PLN20X, numerical aperture NA=0.4). The resulting spots have a waist of 800 nm, characterized by re-imaging the array onto a camera using a higher NA objective (PLN40X, NA=0.6). Distortion in this imaging system limits the accurate characterization to the central $\sim2,000$ sites (Fig. \ref{fig:objective}a). After initial alignment, we observe significantly degraded crosstalk on nearest neighbor sites (approximately $7\times10^{-4}$), which we attribute to aberrations in the microscope objective. However, iterative correction of these aberrations with the LCOS-SLM (applying an rms wavefront correction of $\omega_{\rm sph}=0.3$ waves) suppresses the averaged crosstalk to $1.3\times10^{-4}$, only slightly worse than what was observed with the doublet lens (Fig. \ref{fig:objective}b). Uniform performance is observed across the portion of the array that we can characterize (Fig. \ref{fig:objective}c, d).

We now comment on several aspects of the modulator performance and potential applications. First, in the case of locally addressed gates for quantum computing, the added gate errors from the modulator will depend on the type of gate being implemented. This modulator system is particularly well-suited to gates controlled by light shifts~\cite{labuhn2014single,deleseleuc2017optical,levine2018high}, particularly with nuclear spin qubits~\cite{burgers2022controlling, ma2022universal, lis2023mid, huie2023repetitive, norcia2023mid}, as this approach is extremely robust to intensity fluctuations. In this case, the addressing errors will be predominantly from the finite contrast and crosstalk, and should be at the level of $10^{-4}$. On the other hand, gate implementations involving directly driving an atomic transition with a focused beam are more sensitive to intensity errors~\cite{debnath2016demonstsration,graham2022multiqubit}, though this can be mitigated using robust pulses~\cite{brown2004arbitrarily, merrill2014progress, fromonteil2023, jandura2023optimizing}.

Second, while our demonstration uses a wavelength of 632.8 nm and a modest total power of 2 mW (0.2 $\mu$W/site), the underlying components can operate at wavelengths from 365~nm to beyond 1~$\mu$m, and at power levels up to $100$~W~\cite{hamamatsu_slm, ti_dmd} (10~mW/site), depending on the wavelength and pulse duty cycle. The total power efficiency of the LCOS-SLM and DMD portion of the modulator is approximately 0.13 for a 10,000-spot array. The efficiency of the separate components is approximately 0.37 for the LCOS-SLM, 0.70 for the DMD, and 0.50 for the compensation grating.

Third, we note that the comparison between the performance obtained with an achromatic doublet lens (Figs. \ref{fig:uniform}, \ref{fig:crosstalk}) and a microscope objective (Fig. \ref{fig:objective}) illustrates the role of even very low levels of aberrations on the crosstalk and contrast. To obtain optimal performance with qubits, it will be necessary to implement techniques for in situ characterization of aberrations~\cite{zupancic2016ultra}. In this context, the LCOS-SLM is beneficial for performing fine adjustments of the aberrations.

For smaller arrays or fortuitous combinations of wavelength and DMD blaze angle, the aberrations from misalignment of the DMD with the image plane are small enough that they can be pre-compensated with the LCOS-SLM, avoiding the need for a compensation grating (see Supplementary Information). However, the compensation grating presented here allows the LCOS-SLM to be replaced with other modulators with less flexible wavefront shaping, such as AODs. The combination of AODs and a DMD may provide greater power efficiency when driving gates on a very sparse subset of the entire array, and the ability to selectively switch off sites may enable new avenues for optical tweezer rearrangement.

Finally, we consider possibilities for scaling to larger arrays and realizing faster switching. The 10,000-site array in the present work corresponds to 210,000 diffraction-limited modes, when considering the spacing of $4.6w_0$. This is close to the Nyquist limit for both the LCOS-SLM and the DMD, which have approximately 1 megapixel of resolution. Given the aberration control demonstrated in this work, scaling to larger arrays should be straightforward with higher resolution modulators, or by tiling multiple modulators. The DMD switching speed is limited by both the data transfer rate and the mechanical response time of the pixels. An improvement of one order of magnitude is feasible by using smaller micromirrors and grouping them together electrically to reduce the data rate.

In conclusion, we have demonstrated a new approach to locally controlling gate operations in an array of optically addressed qubits. By leveraging existing optoelectronic devices, we can individually control the intensity of 10,000 sites at a frequency of up to 43 kHz. We comprehensively characterize the intensity and beam size uniformity, as well as the contrast, crosstalk, and dynamic performance. This local gate controller offers a promising and effective solution to scalable quantum computing with optically-addressed qubits.

\begin{acknowledgments}
\emph{Acknowledgements} We gratefully acknowledge Shuo Ma, Genyue Liu, Alex Burgers, Sebastian Horvath, Rajibul Islam, Chung-You Shih and Markus Greiner for helpful conversations, and Adam Kaufman for feedback on a manuscript draft. This work was supported by the Army Research Office (W911NF-1810215), the Office of Naval Research (N00014-20-1-2426, N00014-23-1-2621) and DARPA ONISQ (W911NF-20-10021).
\end{acknowledgments}

\clearpage

\clearpage
\renewcommand{\thefigure}{S\arabic{figure}}
\renewcommand{\theequation}{S\arabic{equation}}
\setcounter{figure}{0}
\setcounter{equation}{0}
\section{Supplementary Information}

\subsection{Experimental apparatus}
\begin{figure}[!htp]
    \centering
    \includegraphics[width= 180 mm]{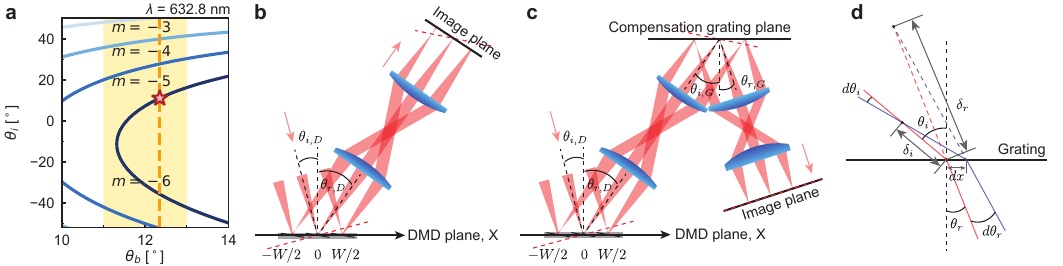}
    \caption{\textbf{Optical system analysis} (a) The incidence angle of the DMD is chosen to satisfy the blazing condition, given the operating wavelength (632.8 nm), the DMD pitch (exactly 13.68~$\mu$m), and blaze angle ($11^\circ$--$13^\circ$ for the TI DLP7000, represented by the yellow-shaded area, but measured to be $12.35^\circ$ for the particular device that we use, represented by the orange dashed line). We operate at the point indicated by the red star. (b) Depiction of system aberrations arising from the tilted optical axes. The focal planes of the imaging system, represented by red dashed lines, do not align with the DMD or image planes. Grey box illustrates the DMD aperture width $W$. (c) Principle of aberration correction using a compensation grating. (d) Depiction of a grating altering the effective focal point within the $xz$ plane. A transmission grating is drawn here for clarity, reducing figure congestion. This concept also extends to reflective gratings.}
    \label{fig:defocus}
\end{figure}

An overview of the experimental apparatus is depicted in Fig.~\ref{fig:intro}a. The primary laser beam is produced by a helium neon laser (632.8 nm), modulated by an AOM to produce pulses. The beam is then coupled into a single-mode fiber (Sch\"after+Kirchhoff PMC-630Si-4.2-NA012-3-APC-500-P), and then back into free space in front of the LCOS-SLM (Hamamatsu x15213-01). The collimator is a Schäfter+Kirchhoff 60FC-L-4-M40L-26, with a focal length of $f=40$ mm to create a beam waist of approximately 4 mm ($1/e^2$ radius). This is chosen to be approximately 60\% of the semi-height of the LCOS-SLM aperture to minimize diffraction from the aperture edges, which can lead to worse optical crosstalk in the image plane. We focus the secondary beams generated by the LCOS-SLM onto the DMD with a $f=400$ mm achromatic doublet lens (Thorlabs ACT508-400-A). The DMD is a DLP7000 chip from Texas Instruments, which has $1024 \times 768$ pixels, with a pitch size of $a = 13.68\,\mu$m, driven by a ViALUX V4395 controller. The specified blaze angle of the DLP7000 is $11^\circ$--$13^\circ$, and a measurement of our specific device using the blazing efficiency reveals an angle of $\theta_{b, D}=12.35^\circ$. The angle of incidence (AOI, $\theta_{i,D}=11.1^\circ$) to the DMD plane is chosen to satisfy a blazing condition (see Fig. \ref{fig:defocus}a). After the DMD, a $M=1:1$ telescope made from $f=200$ mm achromatic lenses relays the modulated beams to a ruled grating (Edmund optics 41-025, 600 grooves/mm, $\theta_{b, G}=13^\circ$) at an AOI of $\theta_{i,G}=36.8^\circ$, to undo defocus and aberrations from the DMD (the basis for these choices is explained in the following section). To characterize the performance of the modulator, the resulting array is imaged onto a camera (FLIR BFS-PGE-200S6C-C) using a $f=300$ mm lens (Thorlabs ACT508-300-A) and a $f=200$ mm objective lens (Thorlabs TTL200-A). For the data in Fig. \ref{fig:dynamics}, the camera is replaced with a photodiode (Thorlabs APD130A). For the data in Fig. \ref{fig:objective}, the 200 mm objective lens is replaced with a microscope objective (Olympus Plan 20X Achromat Objective), to create an array with sub-micron spot sizes and 3.15~$\mu$m spacing. This array is imaged onto the camera using a higher-NA objective (Olympus Plan 40X Achromat Objective).

\subsection{Aberration analysis and correction }

This section discusses the principles governing the use of the DMD with an imaging system with a tilted optical axis. We compute the dominant aberrations (defocus and astigmatism) in the plane of the DMD and in the final image plane. The aberrations in the image plane are significant when generating large spot arrays, but can be corrected using a compensation grating as discussed in the main text. We also discuss an alternative strategy of precompensation with the LCOS-SLM, which may be suitable in some cases. The aberrations in the DMD plane are not a significant problem for the parameters of this work, but could limit the array size and crosstalk in certain cases including scaling to larger DMDs. In this case, we propose a solution that combines a compensation grating and LCOS-SLM precompensation.

The aberrations are related to the tilted optical axes shown in Fig.~\ref{fig:defocus}b. The angle of incidence on the DMD, $\theta_{i,D}$, is chosen to satisfy the blazing condition. As illustrated in Fig. \ref{fig:defocus}a, for each order $m \in \mathbb{Z}$, we have a pair of potential values for $\theta_{i, D}$, given by:
\begin{equation}
  \theta_{i, D} = \pm \arccos \left( \frac{- m \lambda}{2 d \sin \theta_b}
  \right) - \theta_b.
\end{equation}
 Here we follow the convention that $m=0$ corresponds to the specular reflection [$\theta_i=\theta_r(m)$]. We prefer an $m$ such that $\arccos \left( \frac{- m \lambda}{2 d \sin \theta_b} \right) - \theta_b > 0$, which ensures that the incident and exiting beams are well-separated. Additionally, it is crucial that $|\theta_{i,D}|$ remains as small as possible in order to minimize optical power leakage through the gaps between tilted micromirrors, thereby achieving high power efficiency. A configuration satisfying $\theta_{i,D} = \arccos \left( \frac{- m \lambda}{2 d \sin \theta_b} \right) - \theta_b>0$ and $\theta_{r,D}=\arccos \left( \frac{- m \lambda}{2 d \sin \theta_b} \right) + \theta_b>0$ is depicted in Fig. \ref{fig:defocus}b.

We first compute the aberrations in the image plane, which is the primary limitation in our current configuration. The grating nature of the DMD results in a misalignment between the image plane and the focal plane of the imaging system (Fig. \ref{fig:defocus}b). The defocus relative to the Rayleigh length remains the same while beams pass through the telecope between the DMD and image planes. Therefore, by examining the defocus as beams leave the DMD plane, we can infer the defocus in the image plane. After beams exit the DMD plane, the defocus for rays in $xz$ and $yz$ planes increases linearly with the displacement $X$ from the center of the DMD aperture, as:
\begin{align}
    \delta_{x, I} &=  (\sin \theta_{i, D} \beta^2_D - \sin \theta_{r, D}) X,\\
    \delta_{y, I} &=  (\sin \theta_{i, D} - \sin \theta_{r, D}) X,
  \end{align}
where $\beta_D = \cos \theta_{r, D} / \cos \theta_{i, D}$. To achieve diffraction-limited performance (\emph{i.e.}, peak-to-valley wavefront error below $\lambda / 10$), the defocus must be less than $0.3 z_R$ , where $z_R$ is the Rayleigh length in the DMD plane. For the DMD parameters and beam width (20.4~$\mu$m in the DMD plane) in this work, this condition is only satisfied for $|X| \lesssim W/12$ (where $W$ is the width of the DMD aperture).

The aberrations in the image plane can be corrected by adding a telescope ($M$) and a compensation grating ($\beta_G = \cos \theta_{r, G} / \cos \theta_{i, G}$) as illustrated in Fig. \ref{fig:defocus}c. In this configuration, the defocus in the image plane is:
\begin{align}
  \delta_{x, I}' &= M \left[ M \sin \theta_{i, D} \beta_D^2 \beta_G^2 - M
  \sin \theta_{r, D} \beta_G^2 + \frac{\cos \theta_{r, D}}{\cos \theta_{i, G}}
  (\sin \theta_{i, G} \beta^2_G - \sin \theta_{r, G}) \right] X, \\
  \delta_{y, I}' &= M \left[ M \sin \theta_{i, D} - M \sin \theta_{r, D}
  + \frac{\cos \theta_{r, D}}{\cos \theta_{i, G}} (\sin \theta_{i, G} - \sin
  \theta_{r, G}) \right]X .
\end{align}

These slopes can be simultaneously zeroed by adjusting $M$ and $\beta_G$. However, given the discrete options for $M$ and $\beta_G$ using stock optical components, we again consider the condition that the defocus should be less than $0.3 z_R$ in the image plane. Using a telescope with $M=1$ and a grating with a groove density of 600 grooves/mm, we determined a pair of incident and reflecting angles, $\theta_{i,G} = 36.8^\circ$ and $\theta_{r,G}=12.7^\circ$, respectively. The defocus in the image plane is less than $0.03 z_R$ across the full DMD aperture (\emph{i.e}, for $|X| \leq W/2$).

We now consider the defocus in the DMD plane. While defocused spots in the DMD plane do not necessarily impair the spot quality in the final image plane, it can lead to crosstalk if the spots begin to overlap on the DMD.
The defocus of the spots in the DMD plane is given simply by
\begin{equation}
    \delta_{D} = \sin \theta_{i, D} X.
\end{equation}
To avoid overlapping spots for an array with a spacing of $A w_0$ ($A > 2$),  the defocus must satisfy $|X \sin \theta_{i, D}|< \varepsilon_D z_R = \varepsilon_D \frac{\pi w_0^2}{\lambda}$, where $\varepsilon_D = \sqrt{ \frac{A^2}{4} - 1}$ ($\varepsilon_D = 2.1$ when $A = 4.6$). Given the wavelength, beam sizes, and DMD parameters utilized in this work, the defocus at the edge of the DMD aperture is $\delta_{D}(W/2) = 0.8 z_R$, leading to a minimum spacing of $A_{\min} = 2.6$. This is about half of the spacing used in this work, so defocus in the DMD plane does not present a limitation.

It is also possible to use the wavefront correction capabilities of the LCOS-SLM to pre-compensate the aberrations induced by the DMD, to achieve a uniform spot array in the image plane without a compensation grating. This will necessarily exacerbate the aberrations in the DMD plane, but as noted above, these are not a significant constraint for the array parameters presented in this work. In this case, it is preferable to select an angle of incidence $|\theta_{i,D}|=|-\arccos \left[{- m \lambda}/{(2 d \sin \theta_b)} \right] - \theta_b|$, ensuring a minimal exiting angle $|\theta_{r,D}|=|\arccos \left[{- m \lambda}/{(2 d \sin \theta_b)} \right] - \theta_b|$. For the DMD in this work, this would favor reversing the input and output paths.

Lastly, we note that in future devices where it is desirable to scale to even larger number of spots (\emph{i.e.}, by decreasing the spacing or using DMDs with larger numbers of pixels), the DMD-plane aberrations may become problematic. This can be addressed by using the LCOS-SLM pre-compensation to remove the aberrations in this plane, at the expense of increasing the aberrations at the image plane. However, the image plane aberrations can subsequently be corrected with a telescope and compensation grating, as discussed above.

\subsection{Grating-induced astigmatism}

Continuing with the assumption that the incident and exiting beams reside in the $xz$ plane as discussed previously, the distance between the focal point and the grating plane is designated as $\delta_i = \delta_{x,i} = \delta_{y,i}$, as illustrated in Fig. \ref{fig:defocus}d. From the standpoint of geometrical optics, the focal point is identified as the intersection of plane waves with varying angles of incidence. A large beam waist, $w_0>a\gg\lambda$, guarantees a slight divergence of these angles. Hence, an incidence angle can be written as:
\begin{align}
    \theta_i (k_x) = \theta_i (0) + \arctan (k_x / k_z) \approx \theta_i (0) + k_x / k_z,
\end{align}
where $k_x$ and $k_z$ represent the wave vectors defining the direction of propagation. The exit angle is determined by the grating equation, $\sin (\theta_r) - \sin (\theta_i) = m
\lambda / d$. By taking derivatives on both sides, we have
\begin{align}
    \frac{d \theta_r}{d \theta_i} = \frac{\cos(\theta_i)}{\cos (\theta_r)},
    \label{eq:angle_relation}
\end{align}
meaning, after passing the grating, the new $x$ wavevector becomes $k_x' = \frac{\cos (\theta_i)}{\cos (\theta_r)} k_x$, whereas the new $y$ wavevector remains $k_y' = k_y$, since the grating vector is along the $x$ direction.

In Fig. \ref{fig:defocus}d, the intersection of two plane waves (shown in red and blue) designate the focal point of the incident beam. The divergence angle of the two wave planes prior to the grating $d\theta_i\ll1$, hence, the distance in the grating plane is approximately $dx=\delta_{x,i}d\theta_i/\cos\theta_i$. The divergence angle of the exiting plane waves is also small ($d\theta_r=d\theta_i\cos\theta_i/\cos\theta_r \ll 1$). Consequently, we establish the effective focal point locations for rays in $xz$ and $yz$ planes post-diffraction, respectively:
\begin{align}
    \delta_{x,r} &= \frac{dx}{\cos\theta_r}/d\theta_r= (\cos\theta_r/\cos\theta_i)^2\delta_{i}, \\
    \delta_{y,r} &= \delta_{i}.
\end{align}
The occurrence of astigmatism is due to $\delta_{x,r}\neq\delta_{y,r}$.
It should be noted that astigmatism can also be deduced by decomposing a Gaussian beam into plane waves via a Fourier transform, applying Eq. \ref{eq:angle_relation} to plane waves, and then conducting an inverse Fourier transform.

\subsection{Numerical simulation of on/off contrast ratio}

\begin{figure}[!htp]
    \centering
    \includegraphics[width=115mm]{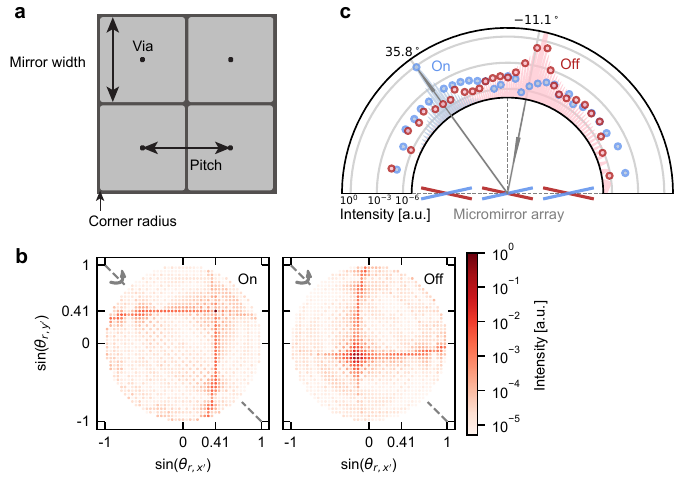}
    \caption{{\bf Amplitude modulation via controlling blazing condition of DMD} (a) The physical dimensions of the micromirror array used in the FDTD simulation. (b) Numerically simulated intensity distribution on all diffraction orders using FDTD method. Diagonal grey dashed lines mark the direction of the micromirror hinge. Dots at $\sin(\theta_{r,x'})=\sin(\theta_{r,y'})=0.414$ represent the captured diffraction order. The simulated on/off contrast for the order is $5.9\times10^{-6}$. (c) Comparison between the theoretical perfect 2D grating power envelope and numerically simulated intensities via FDTD method for the diagonal diffraction orders ($m_{x'}=m_{y'}$). The blue-shaded area represents the power envelope when the DMD is in the ``on'' state, while the red-shaded area corresponds to the ``off'' state. The blue and red dots indicate the numerically simulated intensities of diagonal orders in ``on'' and ``off'' states, respectively.
}
    \label{fig:fdtd}
\end{figure}

In this section, we present analytic and numerical models to understand the experimentally measured on/off contrast. The diffraction pattern produced by a single micromirror on the DMD manifests an intensity envelope that peaks in the direction of the micromirror's specular reflection. The interference of the reflection from many mirrors further discretizes the diffraction envelope into separate peaks marked by orders in two directions, $m_{x'}$ and $m_{y'}$. Here the $x'$ and the $y'$ axes are aligned to the edges of the square mirrors, and are rotated by 45$^\circ$ from the coordinate frame in Fig.~\ref{fig:defocus}.
The diffraction envelopes peak along the $x'$ and the $y'$ axis, respectively, resulting in an enhanced contrast ratio for the diagonal orders ($m_{x'}=m_{y'}=m$). The normalized optical power for a diagonal order $m$ in a perfect 2D grating is given by the equation
\begin{align}
    I(m) = \frac{\tmop{sinc}^4 \left[ \frac{\pi h}{\sqrt{2} \lambda} (\sin
   \theta_r (m) - \sin (\theta_i + 2 \theta_b)) \right]}{\sum_{m_{x'}',
   m_{y'}'} \tmop{sinc}^2 \left[ \frac{\pi h}{\lambda} \left( \sin \theta_{r,
   x'} (m_{x'}') - \frac{\sin (\theta_i + 2 \theta_b)}{\sqrt{2}} \right)
   \right] \cdot \tmop{sinc}^2 \left[ \frac{\pi h}{\lambda} \left( \sin
   \theta_{r, y'} (m_{y'}') - \frac{\sin (\theta_i + 2 \theta_b)}{\sqrt{2}}
   \right) \right]}.
   \label{eq:sinc}
\end{align}
where $h$ is the width of a micromirror and $\theta_{r,x'}$ ($\theta_{r,y'}$) is the exiting angle in $x'z$ ($y'z$) plane. For the value $m=6$ used in this work, Eq.~\ref{eq:sinc} predicts a contrast of $5\times10^{-7}$ for a plane wave input.

To understand the role of non-ideal effects in a realistic DMD, we deploy an FDTD simulation \cite{taflove2005computational} to estimate the contrast, again with plane wave illumination. The physical dimensions of the model (Fig. \ref{fig:fdtd}a) are the pitch size $a=13.68~\mu$m, the micromirror width 13.14~$\mu$m, the via width  0.75~$\mu$m, the via depth 1.75 $\mu$m, and the micromirror tilt 12.35$^\circ$. Both the corner radius and the fillet radius are set at 0.4~$\mu$m~\cite{ti_dmd, piotrowski2022optical}. While the DMD mirrors have an aluminum coating, the simulation uses a perfect electric conductor boundary to reduce the simulation time. Fig. \ref{fig:fdtd}b shows the simulation results when the DMD is in the ``on" and ``off" states. When the micromirrors are in the on state, approximately 73\% of the total incident power is channeled into the desired order. Conversely, when the micromirrors are in the off state, the desired order is significantly distanced from the envelope's center, yielding an intensity $4.3\times10^{-6}$ times that of the incident intensity (\emph{i.e.}, an extinction ratio of $5.9\times10^{-6}$). This is relatively close to the experimental value. The remaining discrepancy may be attributed to device non-idealities not captured into the model, or to the finite beam waist of the input beam.

\bibliography{refs}

\end{document}